 \documentclass[final,3p,times]{elsarticle}

\usepackage{journalnames} 
 \usepackage{graphicx}
 \usepackage{epstopdf}

\usepackage{amssymb}
\usepackage{amsthm}
\def\astrobj#1{#1}

\journal{New Astronomy}

\begin{document}

\begin{frontmatter}



\title{Is the \astrobj{Sgr dSph} a dark matter dominated system?}

\author[UN]{C.A. ˜Mart\'inez-Barbosa\corref{cor1}\fnref{L}}
\ead{camartinezba@unal.edu.co, carmen@strw.leidenuniv.nl}

\author[UN]{R.A. ˜Casas\corref{cor2}}
\ead{racasasm@unal.edu.co}

\cortext[cor1]{Principal corresponding author}
\cortext[cor2]{corresponding author}

\address[UN]{Universidad Nacional de Colombia, \\Cra 30 No. 45-03 Bogot\'a, D.C. Colombia}
\address[L]{Leiden observatory, Leiden university, P.O. Box 9513, NL-2300 RA Leiden, \\The Netherlands}

\begin{abstract}
We study the evolution of  possible progenitors of \astrobj{Sgr dSph} using several numerical N-body simulations of different dwarf spheroidal galaxies both with and without dark matter, as they  orbit the Milky Way. The barionic and dark components of the dwarfs  were made obeying a Plummer and NFW potentials of $10^6$ particles respectively. The Milky Way was modeled like a tree-component rigid potential and the simulations were performed using  a modified Gadget-2 code.  We found that none of the simulated galaxies without dark matter reproduced the physical properties  observed in \astrobj{Sgr dSph}, suggesting that, at the beginning of its evolution, \astrobj{Sgr dSph} might have been immersed in a dark matter halo. \quad \\
The simulations of progenitors immersed in dark matter halos suggest that \astrobj{Sgr dSph} at its beginning might have been an extended system, i.e.  its Plummer radius could have had a value approximated to 1.2 kpc or higher; furthermore, this galaxy could have been immersed in a dark halo  with a mass higher than  $10^{8}$  $M_{\odot}$. These results are important for the construction of a model of the formation of \astrobj{Sgr dSph}.

\end{abstract}


\end{frontmatter}

\section{Introduction}
\label{intro}

Since its discovery in 1994 by  \citet{ibata95}, the \astrobj{Sgr dSph} dwarf galaxy  has been object of febrile research; but only recently, thanks to the new data provided by 2MASS and SDSS, its tidal tails have been studied in  great detail 
\citep{newberg02,  helmi03, law03, majewsky03a,  fellhauer06, belokurov06, yanny09, watkins09}. The Sgr dSph is very important because it gives us a better understanding about the formation process of the Milky Way and many clues about the shape of its dark matter halo.\\

\astrobj{Sgr dSph} tidal tails are the biggest contributions to the galactic halo \citep{giuffrida09} at a radius above 50 kpc \citep{keller08}. Many studies have found tidally stripped material associated to this dwarf with an age between 3 and 6.5 Gyrs \citep{belokurov06}; others, suggest that the age of the older debris is between 2 and 4 Gyrs \citep{helmi04}; furthermore, arc substructures have been found through the galactic south hemisphere at a distance of approximately 40 kpc \citep{majewsky03a}; as well as  filaments at a distance of 45 kpc \citep{keller08, martinez04}; $46\pm12$ kpc \citep{martinez01} and $62\pm6$ kpc \citep{martinez04} to the galactic center. These data suggest that \astrobj{Sgr dSph} has been disrupted since about 6 Gyrs. \\

The nearest \astrobj{Sgr dSph} tidal tails have been found at 18 kpc from the galactic center. These debris belong to filaments that emerged during the last pericenter passage of the satellite, approximately 0.75 Gyrs ago \citep{dinescu02}. The most recent discoveries about the structure of the tidal tails of the Sgr dSph, were done by \citet{belokurov06} and \citet{yanny09} who found two branches of the \astrobj{Sgr dSph} stream -the leading and trailing tidal tails- in an area around the north galactic cap.\\

\begin{table}
\centering
  \begin{minipage}{70mm}
  \caption{Physical properties observed in \astrobj{Sgr dSph} by several authors. The nomenclature is referred to: *\cite{mateo98} , ** \cite{ibata98}, *** \cite{gomez99},  $^{+}$ \cite{law05}}
  \label{t1}
  \centering
\vspace{1mm}
  \begin{tabular}{c  c}
  \hline
$r_{1/2}$  & 0.55 kpc*** \\[2mm]
$\sigma_0$ & $11.4\pm 7$ km/s ** \\ [2mm]                 
$\mu_{0,v}$ & 25.4 mag/arcsec$^2$ ***\\
                    & $2.47\times 10^6$ $L_{\odot}$/kpc$^2$ \\ [2mm]
M/L           & 10 * \\
                 & 14-16 $^{+}$ \\ [2mm]
\hline
\end{tabular}
\end{minipage}
\end{table}

\astrobj{Sgr dSph} main body is located at $16\pm2$ kpc from the galactic center \citep{johnston95, irwin95, ibata97, ibata99, gomez99}, at a distance of $24\pm2$ kpc  from the sun \citep{johnston95, alard96, ibata98} with  $l=5^o$, $b=-14.5^o$ \citep{ibata98, belokurov06}.  This dwarf covers a vast region on the sky, $15^o\times 7^o$ \citep{giuffrida09}.  Numerical simulations suggest that it is not symmetric; that is, the main body of \astrobj{Sgr dSph} is a prolate object with axis ratios $\approx$ 3:1:1 \citep{ibata97, ibata99}. \\

The central velocity dispersion of the main body of \astrobj{Sgr dSph} is $11.4\pm 0.7$ km/s \citep{ibata97, ibata98}. Kinematical studies suggest that \astrobj{Sgr dSph} moves to the north with transversal velocity of $250\pm90$ km/s \citep{irwin96, ibata98}. The radial velocity of the satellite is 171 km/s \citep{irwin96, johnston99, gomez99, majewsky03c, law03, law05}. \\

Studies realized by \citet{law03} and \citet{majewsky03c} indicate that the mass bound to the main body of Sgr dSph is $\approx 3\times10^8$ $M_{\odot}$. A better estimate was done by \citet{law10}, who conclude that  $M=2.5^{+1.3}_{-1}\times 10^8$ $M_{\odot}$; with this range of masses, the absolute visual magnitude of the satellite is  $M_v= -13.64$; its total luminosity, $L_T=2.4\times 10^7$$L_{\odot}$ and its mass to light ratio, $M/L=10^{+6}_{-4}$ $M_{\odot}/L_{\odot}$. Table \ref{t1} shows some physical properties of this dwarf observed by other authors. $r_{1/2}$ is the half light radius; $\sigma_0$ is the central line of sight velocity dispersion; $\mu_0$ is the central surface brightness and M/L is the mass to light ratio.\\

The aim of this paper is to show possible features of the progenitor of \astrobj{Sgr dSph} and an estimate of the mass of its initial dark matter halo using N-body simulations.  We present our model in section \ref{model}; the results obtained by simulating several possible progenitors of \astrobj{Sgr dSph} without dark matter are presented in section \ref{BM}.  Some of these galaxies were simulated in different contents of dark matter. These results are shown in Section \ref{DM}. Finally we summarize some features of both the progenitor of \astrobj{Sgr dSph} and its initial dark halo.

\section{Model}
\label{model}

\subsection{Initial conditions}
\label{IC}

\begin{figure}
\centering
\includegraphics[width=7cm, height=5cm]{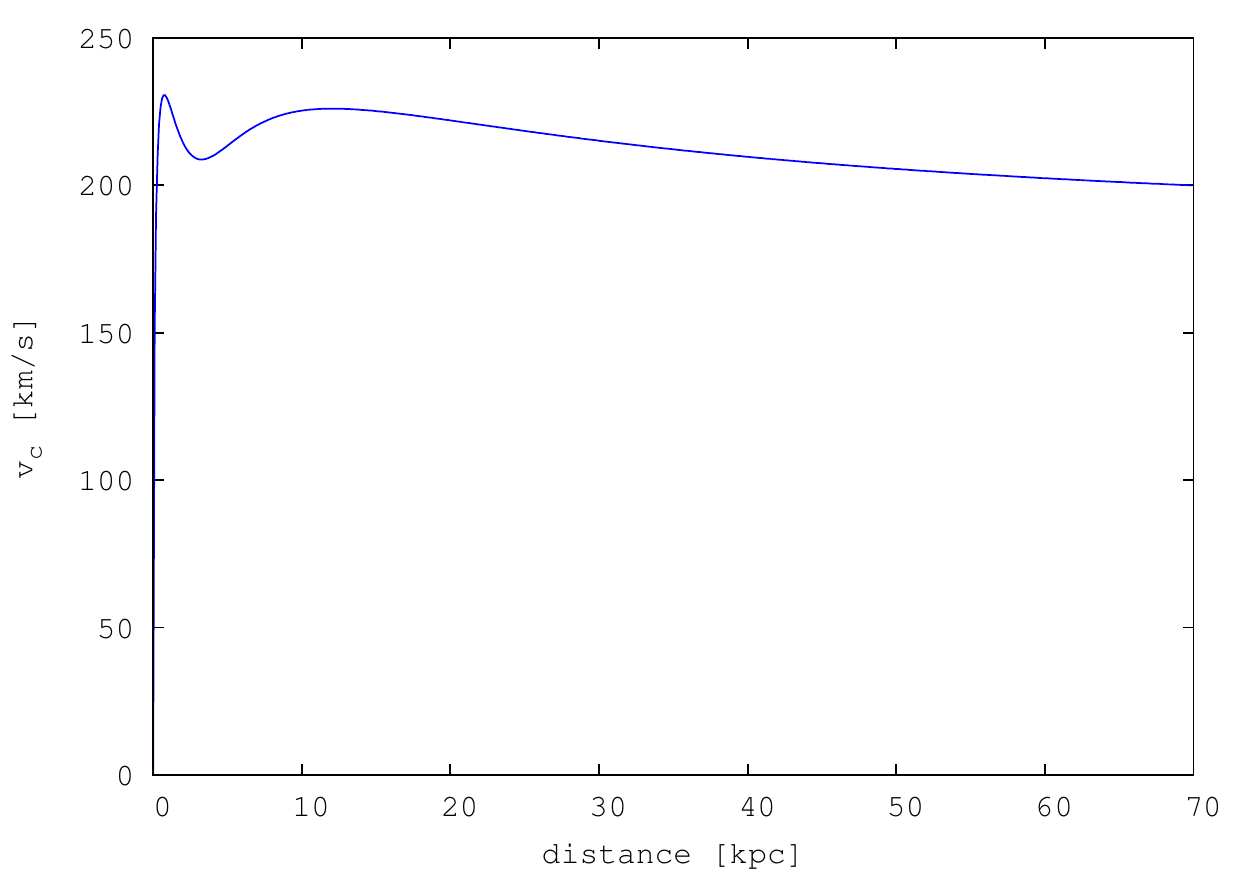}\\
\caption{Rotation curve of the Milky Way using the parameters of \citet{johnston95}.  }
\label{f1}
\end{figure}

There have been several efforts to unravel the real shape of the dark matter halo of The Milky Way. Some studies done by \citet{ibata01} and \citet{majewsky03a} suggest that its shape must be close to spherical; \citet{helmi04} argues that the velocities of the leading tidal tail favor a prolate dark matter halo; nevertheless, the over-densities  located at 90 kpc from the galactic center found by \citet{newberg03} suggest that the dark halo is nearly spherical or slightly oblate \citep{yanny09}. The mystery seems to be solved taking into account the bifurcations in the tidal tails observed by \citet{belokurov06} and \citet{yanny09}. \citet{fellhauer06} found that, if the potential of the dark halo were exactly spherical, the debris of \astrobj{Sgr dSph} would lie in a single plane and no bifurcation would exist; furthermore, If the halo is too oblate or too prolate, then debris are scattered over a wide range of locations. Furthermore, if the simulated satellite has a considerable mass (more or equal to $7.5\times 10^8$ $M_{\odot}$), the bifurcations blurs. \citet{penarrubia10} found that bifurcations in the leading tail of \astrobj{Sgr dSph} stream naturally appear  in models where the dwarf is represented by an exponential disk and if it is misaligned with respect to the orbital plane; furthermore, \citet{law10} reproduced the majority of major constraints of the angular positions, distances and radial velocities of the young tidal debris streams, introducing a non axisymmetric potential in the form of a nearly oblate dark matter halo.  With these results, both studies make a realistic picture of \astrobj{Sgr dSph}.\\  
 
Since we aim to find out if \astrobj{Sgr dSph} must have dark matter for its evolution and what was its initial content, we will only focus on reproducing  the physical properties observed in the main body, not in reproducing the bifurcations of the tidal tails. To do that, the Milky Way is modeled as a three-component rigid potential, in which the disk is represented by a Miyamoto \& Nagai potential \citep{miyamoto75}, the spheroid by a Hernquist potential \citep{hernquist90}  and the dark halo by a spherical logarithmic potential \citep{johnston99}:

\begin{large}
 \begin{eqnarray}
\phi_{d} & = -\frac{GM_{d}}{\sqrt{R^2 + (a+\sqrt{z^2+b^2})^2}} \\
\phi_{s}  &=  -\frac{GM_{s}}{r+c}\\
\phi_{h} &=  \nu^2_{h}\ln{(r^2+d^2)}.
\end{eqnarray}
\end{large}

Where the disk mass $M_{d}=1\times10^{11}$ $M_{\odot}$; the spheroid mass $M_{s}=3.4\times10^{10}$ $M_{\odot}$; the dark halo circular velocity $\nu_h=128$ km/s; and the lenght scales related to each components of the Milky Way: a=6.5 kpc; b=0.26 kpc; c=0.7 kpc; d=12 kpc. These parameters provide a nearly flat rotation curve between 1 and 30 kpc and a disk scale height of 0.2 kpc for our galaxy as can be seen in figure \ref{f1}. \\

If the Milky Way is modeled as a rigid potential, we can isolate the effects of tidal interactions on the internal structure and kinematics of the dwarf \citep{klimentowski09}; although, with this approach, we neglect the effects of dynamical friction, the evolution of the potential of the host galaxy and interactions between dwarfs.\\

With the purpose of establishing the initial conditions that could have the progenitor of \astrobj{Sgr dSph}, a test particle was evolved under the action of the rigid potential,  with the current position and radial velocity of \astrobj{Sgr dSph}; that is 16 kpc and 171 km/s respectively. Since the tangential velocity can not be measured, it is taken as a free parameter; we  varied it from -325 to 325 km/s because with these values the apocenter of \astrobj{Sgr dSph} will be between 40  and 60 kpc, which is in agreement with the tidal tails found by  \citet{martinez04}.  Using numerical simulations, \citet {martinez04}  found that these tidal tails were disrupted from the satellite approximately 6 Gyrs ago. This period of time is very advantageous since  some galactic evolution and dynamical friction effects  can be neglected. \citep{martinez04}.\\ 

Higher values in the tangential velocity will cause the satellite to move farther from the galactic center; hence  dynamical friction effects for the evolution of  \astrobj{Sgr dSph} should be taken into account. Thus, we will assume that \astrobj{Sgr dSph} has been orbiting the Milky Way for a period of time smaller than 7 Gyr.\\

For each tangential velocity value, we obtained a Rosette-like orbit for the motion of the test particle; from each path, we computed its apogalactic distance, the tangential velocity at this point and its pericenter distance. Each of these results could become a set of possible initial conditions for the progenitor of \astrobj{Sgr dSph}. In order to establish which of these values reproduce its current galactocentric position and radial velocity,  the target particle was located initially at each apogalactic distance in the Y axis  with an initial velocity given by the velocity at that point. This particle was evolved  under the influence of the rigid potential.  The values which  reproduce the current position and radial velocity of \astrobj{Sgr dSph}, are  shown in table \ref{t2} .\\

\begin{table}
 \centering
 \begin{minipage}{65mm}
  \caption{Initial conditions that reproduce, after a time, the actual galactocentric position and radial velocity of \astrobj{Sgr dSph}. A is the apocenter; P the pericenter; V is the tangential velocity of the particle at the apocenter and $\epsilon_0$, the excentricity.}
   \label{t2}
  \centering
  \begin{tabular}{@{} c c c c}
  \hline
 \textbf{A (kpc)} &\textbf{P (kpc)} & \textbf{V (km/s)} & $\epsilon_0$ \\ \hline \hline
44                     &  12                      &   -104                   &  0.56  \\ 
45                     &  12                      &   -102                  & 0.56    \\  
46                     &  12                      &   -101                  &0.56   \\ 
47                     &  12                      &   -100                  & 0.57  \\ 
48                     &  12                      &    97                    & 0.59 \\ 
49                     &   12                     &    96                    & 0.59  \\  
50                     &  12                      &  95                      &0.60  \\ 
51                     &  12                      &  94                      & 0.60  \\ 
52                     &  12                      &  93                      & 0.61  \\ 
53                     &  13                      & -90                      &0.60 \\ 
54                     &  13                      &  -90                    & 0.60  \\ 
60                     &  13                      & -86                      &  0.63   \\
 \hline
\end{tabular}
\end{minipage}
\end{table}

As we can see in table \ref{t2}, these results are in good agreement with studies made by  \citet{jiang00}\footnote{In this study they found two consistent solutions for the Srg past: one which it has an initial dark halo that caused to sink inwards under the influence of dynamical friction and another where Sgr was moved at all times on the same short period orbit.}; \citet{ibata98}; \citet{law03, law05} and \citet{majewsky03c}. \astrobj{Sgr dSph} apocenter is between 44 and 60 kpc and its pericenter, between 12 and 13 kpc. The orbit period is 0.7 Gyrs.  We also found that the present tangential velocity of the dwarf should be  $298\pm 27$ km/s. \\

\subsection{Dynamical friction}

Since we modeled the Milky Way as a three component rigid potential, we can not simulate massive systems. We used the equation of \citet{binney} \footnote{Despite of this equation is valid only when the host galaxy is modeled as an isothermal sphere, we will use it because most of the orbit of \astrobj{Sgr dSph} lies in the galactic halo.} \ To estimate the maximum mass that we can use in our simulations: 

\begin{equation}
M =\left(\frac{ 2.64 \times 10^{11} }{\log \Lambda}\right)\left(\frac{r}{2 kpc}\right)^2\left(\frac{v_c}{250km/s}\right) \left(\frac{10^{6}yr}{t_{fall}}\right) M_{\odot}
\label{fric}
\end{equation}

Where Coulomb logarithm is $\log \Lambda$=10 for the case when a globular cluster is decaying into the center of the Milky Way,  r and $v_c$ are the initial position and the circular velocity at  this point of the satellite respectively\footnote{This calculation assumes the cluster maintains a circular velocity as it spirals to the center of the Milky Way.}. For our simulations  we take  r=44-60 kpc. From our model of the rigid potential, the circular speed for each position is $v_{c}$= 207.802 - 202.345 km/s respectively.\\

Since our simulation time is $t_{sim} \approx 7$  Gyr, the falling time or $t_{fall}$ should be much higher, we chose $T_{fall}= 10 Gyr$.\\

With those data, we found two maximum values for the initial mass of the system. If $r= 44$ kpc, $M_{max}= 1.06 \times 10^{9}$ $M_{\odot}$; when $r= 60$ kpc, $M_{max}= 1.92 \times 10^{9} $ $M_{\odot}$.

\subsection{Satellites and its  dark halo}
\label{satellite}

Although the representation of \astrobj{Sgr dSph} as a disk, reproduces a realistic picture for its evolution, there is no evidence yet of significant rotation in the main body of the satellite \citep{law10}; that is why the progenitor of \astrobj{Sgr dSph} was modeled as  a Plummer sphere \citep{plummer11} whose potential and density are described respectively by:

\begin{eqnarray}
\phi &=& -\frac{GM_{sat }}{\sqrt{r^2+r^2_0}}\\
\rho &=& \frac{3M_{sat}}{4\pi r_0^3}\frac{1}{\left[ 1+(r/r_0)  \right]^{5/2}}
\end{eqnarray}

where $M_{sat}$ is the initial satellite mass and $r_{0}$ its Plummer radius. We used the Aarseth algorithm \citep{aarseth74, aarseth03} in order to construct the Plummer sphere. The initial mass of the satellites was varied from $5\times10^8$ to $(1-1.9)\times10^9$ $M_{\odot}$, since the present mass of \astrobj{Sgr dSph} is (2.8 - 3.7)$\times10^8$ $M_{\odot}$ \citep{law05}. The upper limit for the initial mass has been chosen to avoid dynamical friction effects as it was shown in the last section. The Plummer radii used were 0.3,0.5,0.6 and 1.2 kpc. \\

In the case of satellites with a dark matter component, the dark matter halo  was modeled using the method of distribution functions proposed by \citet{mashchenko05} which creates a dark halo as a spherical distribution of particles that describes a Navarro-Frenk \& White (NFW) density distribution. The stellar component for these systems was modelled as a Plummer sphere. Since this method provides a distribution function consistent only with dark matter,  we virialized the system composed by the Plummer sphere and the NFW halo with Gadget2 \citep{springel05} until it reaches stability. Again, the maximum initial mass used for simulating systems with barionic and dark matter components  was $(1-1.9) \times 10^9$ $M_{\odot}$ to avoid dynamical friction effects.\\

The simulations of the evolution of the satellites with only barionic matter  and systems of stellar and dark matter components, were performed using a modification of the Gadget-2 code \citep{springel05}. On it, we inserted the rigid potential described in section \ref{IC} to model our galaxy. All the systems were located initially at 44 and 60 kpc from the galactic center, with velocities as given in table \ref{t2}. The softenings lenght used in our simulations were: 0.015 and 0.01 for the halo and stars respectively.

\section{Galaxies without dark matter}
\label{BM}

All the satellites with only barionic matter were virialized during 1 Gyr using the Gadget-2 code in order to stabilize the systems; afterwards, we located them at 44 and 60 kpc from the galactic center with the respective velocities given in table \ref{t2}. Despite of we are not taking into account dynamical friction effects, the current position and radial velocity of \astrobj{Sgr dSph} are reproduced in our N body simulations.\\

\begin{table}
 \centering
 \begin{minipage}{140mm}
  \caption{Physical properties of simulated satellites when they are located initially at 44 kpc from the galactic center. Both Initial mass, $M_{sat}$ and  mass obtained at time t, are given in $M_{\odot}$;  the half light radius $r_{1/2}$ and Plummer radius $r_0$ are in kpc; time when \astrobj{Sgr dSph} is located at the current galactocentric position $t$, is in Gyr; the central surface brightness $\mu_0$ is measured in  $L_{\odot}$/$kpc^2$;  the units of velocity dispersion and mass to light ratio are km/s and $M_{\odot}$/$L_{\odot}$ respectively.}
   \label{t3}
   \vspace{2mm}
  \centering
  \begin{tabular}{@{} c c c c c c c c}
  \hline
 $M_{sat}$ & $r_0$ &\textbf{t} & \textbf{Mass} & $r_{1/2}$ & $\mu_0$ & $\sigma_0$ &\textbf{M/L} \\ \hline \hline
 $1\times 10^9$ & 0.6 & 0.9 & $1\times 10^9$ & 0.18 & $8.18\times 10^8$ & 33.17 & 2.45  \\ [1.5mm] 

$9\times 10^8$ & 0.6 & 0.9 & $7.2 \times10^8$ & 0.18 & $7.24\times 10^8$ & 31.16 & 2.43 \\ [1.5mm]

$8\times 10^8$ & 0.6 & 0.9 & $6.4 \times10^8$ & 0.18 & $6.51\times 10^8$ & 29.08& 2.38 \\ [1.5mm]  
 
$7\times 10^8$ & 0.5 & 0.9 & $7 \times10^8$ & 0.15 & $7.99\times 10^8$ & 30.39 & 2.43  \\ [1.5mm]  

 $6\times 10^8$ & 0.5 & 0.9 & $6 \times10^8$ & 0.15 & $6.85\times 10^8$ & 28.16 & 2.44  \\  [1.5mm]        
 
 $5\times 10^8$ & 0.5 & 0.9 & $4\times 10^8$ & 0.15 & $5.81\times 10^8$ & 25.39 & 2.38 \\  [1.5mm]      
 
 $5\times 10^8$ & 0.3 & 0.9 & $5\times 10^8$ & 0.10 & $51.54\times 10^9$ & 34.19 & 2.39 \\ [1.5 mm] \hline

$7\times 10^8$ & 0.6 & 0.9 & $5.6 \times10^8$ & 0.18 & $5.33\times 10^8$ & 26.92 & 2.43  \\ [1.5mm]

$6\times 10^8$ & 0.6 & 0.9 & $4.8 \times10^6$ & 0.18 & $4.65\times 10^8$ & 24.74&  2.39  \\[1 mm]
                                       
$5\times 10^8$ & 0.6 & 0.9 & $5 \times10^8$ & 0.17 & $4.07\times 10^8$ & 22.17& 2.29  \\ [1mm]\hline
                                                          
$1\times 10^9$ & 1.2 & 0.9 & $4\times 10^8$ & 0.32 & $1.30\times 10^8$ & 21.65 & 3.71\\ [1mm]

$7\times 10^8$ & 1.2 & 0.9 & $2.1 \times10^8$ & 0.34 & $5.8\times 10^7$ & 17.02 & 4.8  \\ [1mm]

 \hline
\end{tabular}
\end{minipage}
\end{table}

\begin{table}
 \centering
 \begin{minipage}{140mm}
  \caption{Physical properties of simulated satellites when they are located initially at 60 kpc from the galactic center. The notation is the same as table \ref{t3}}
   \label{t4}
   \vspace{2mm}
  \centering
  \begin{tabular}{@{} c c c c c c c c}
  \hline
 $M_{sat}$ & $r_0$ &\textbf{t} & \textbf{Mass} & $r_{1/2}$ & $\mu_0$ & $\sigma_0$ &\textbf{M/L} \\ \hline \hline
$1\times 10^9$ & 0.6 & 1.2 & $1\times 10^9$ & 0.18 & $7.8\times 10^8$ & 33.69 & 2.58 \\ [2mm] 
$6\times 10^8$ & 0.6 & 1.2 & $4.8 \times10^8$ & 0.19 & $4.3\times 10^8$ & 25.11& 2.53  \\ [2mm]
$6\times 10^8$ & 0.5 & 1.2 & $6 \times10^8$ & 0.16 & $6.63\times 10^8$ & 28.5 & 2.62  \\ [1mm] 
                         &       & 3.5 & $4.8 \times10^8$ & 0.15 & $6.61\times 10^8$ & 27.05& 2.41  \\ [2mm]
$5\times 10^8$ & 0.6 & 1.2 & $4 \times10^8$ & 0.18 & $3.81\times 10^8$ & 22.44& 2.38  \\ [2mm]                             
$5\times 10^8$ & 0.5 & 1.2 & $5 \times10^8$ & 0.15 & $5.56\times 10^8$ & 25.6& 2.47  \\ [1mm]
                          &      & 3.5 & $3.5\times 10^8$ & 0.14 & $5.49\times 10^8$ & 23.96 & 2.32 \\     \hline

 \hline
\end{tabular}
\end{minipage}
\end{table}

The current physical properties of \astrobj{Sgr dSph} shown in table \ref{t1} were compared with the same quantities calculated in our satellites in order to find out the features of the progenitor of this dwarf galaxy. Theoretical  half light radius, velocity dispersion, surface brightness and mass to light ratio are determined by the method shown by \citet{klessen97}, in which these parameters are calculated by a terrestrial observer. Tables \ref{t3} and \ref{t4} show the results of our simulations when the initial position of the dwarfs is 44 and 60 kpc respectively. For both tables, $\mathbf{t}$ corresponds to the time at which the simulated satellites are located at the current position of \astrobj{Sgr dSph} and have its observational  galactocentric radial velocity; \textbf{Mass} is the mass of these systems at time $\mathbf{t}$. $r_{1/2}$, $\mu_0$, $\sigma_0$ and $M/L$ are the half brightness radius, surface brightness, velocity dispersion and mass to light ratio respectively measured at time $\mathbf{t}$ as well.\\

Comparing the current observed properties of \astrobj{Sgr dSph} with the results from our sample of simulated galaxies, we can see in tables \ref{t3} and \ref{t4} that most of them have a higher mass at time $t$; except from the extended satellites (those which have $r_0= 1.2$ kpc) and the ones with initial mass of $5\times 10^8$ $M_{\odot}$. This is not surprising since the satellites reproduce the current galactocentric position and radial  velocity of \astrobj{Sgr dSph} at early stages of their evolution.\\

The half light radius, $r_{1/2}$, is lower for all the satellites except again for the extended ones  ($r_0= 1.2$ kpc). These satellites reproduce the measured $r_{1/2}$ of \astrobj{Sgr dSph} at 3 Gyrs. Nevertheless, at this time they  do not reproduce the galactocentric position and radial velocity of \astrobj{Sgr dSph}.\\

The surface brightness and the velocity dispersion are larger than the ones measured for \astrobj{Sgr dSph}.\\

\begin{figure}
\centering
\includegraphics[width=12 cm, height=7 cm]{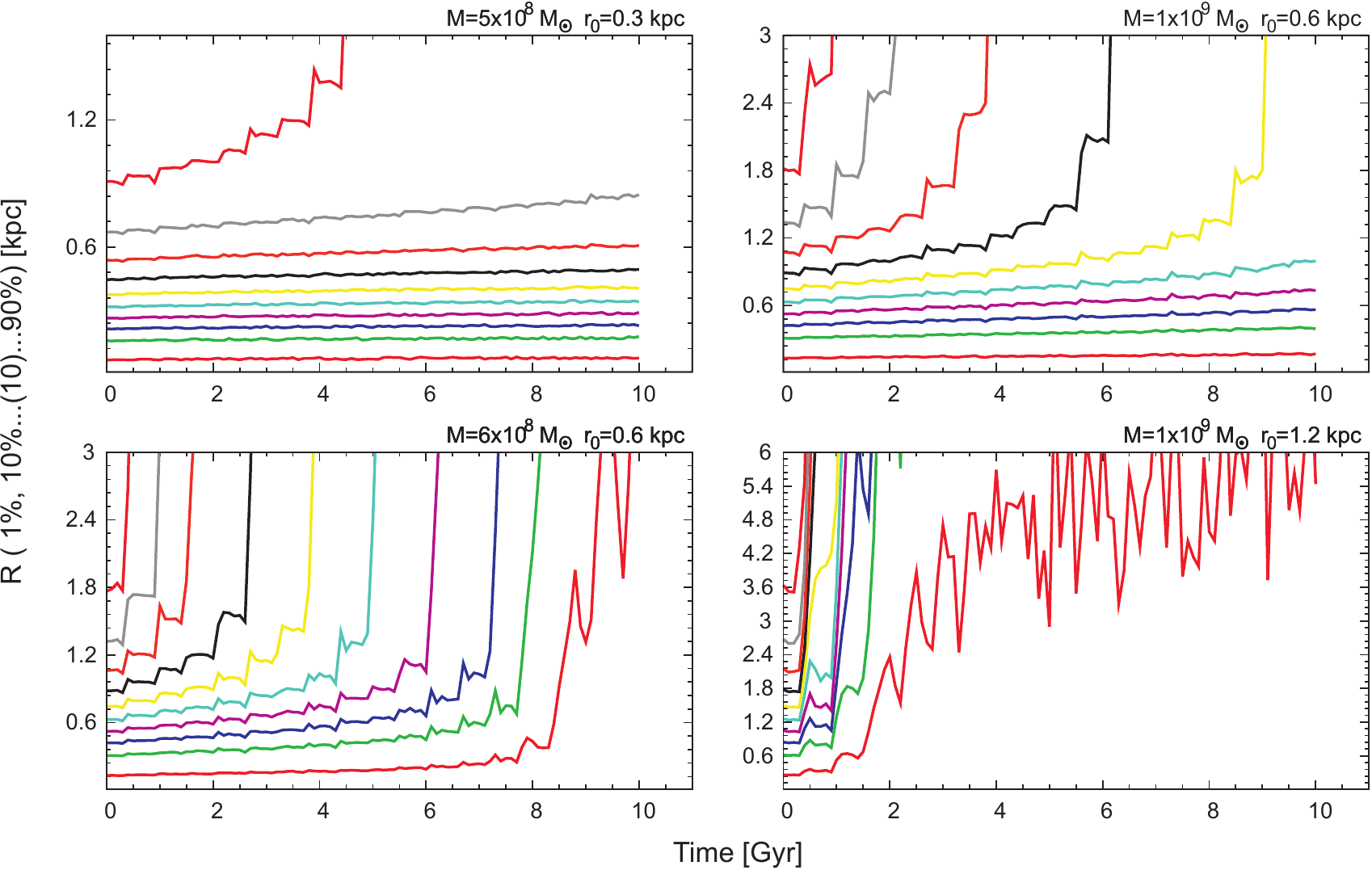}
\caption{Lagrange radii of some simulated satellites when they are located initially at 44 kpc from the galactic center. In all graphics, lines from bottom to top, represent the radius that contains 1\%, 10\%, until 90\% of the initial mass. When the radius which represents 1\% of the initial mass of the galaxy reach the top of each graphic, the satellite is totally disrupted. The maximum number shown in the Y axis, represents the initial virial radius of the satellites, which is calculated using the expression $R=r_0C$ where $r_0$ is the Plummer radius and C=5 is the cutoff taken to construct the Plummer spheres.  }
\label{fig2}
\end{figure}

Figure \ref{fig2} shows the lifetimes for some of the simulated satellites when they are located at 44 kpc from the galactic center. For the case of satellites whose initial mass is very high ($1\times10^9$) $M_{\odot}$ or its Plummer radius is very small (0.3 kpc) more than 30\% of their initial mass survive for 10 Gyrs. It means that they will live for a Hubble time; furthermore, from the same figure,  we can see that these galaxies, between 6 and 7 Gyrs, have masses higher than the current accepted value for \astrobj{Sgr dSph}. It means that these satellites are more stable than the real \astrobj{Sgr dSph}. This  behavior is present in the first galaxies listed in table \ref{t3}. with this evidence, we conclude that the progenitor of \astrobj{Sgr dSph} could not have been a very dense galaxy. In the case of a satellite of initial mass of $6\times 10^8$ $M_{\odot}$ and Plummer radius of 0.6 kpc, this galaxy is totally disrupted before 10 Gyrs. The same behavior is presented in satellites with initial mass of $5\times10^8$ $M_{\odot}$ and Plummer radius of 0.6 kpc.  At 6 Gyr both galaxies reproduce the current mass of \astrobj{Sgr dSph}; nevertheless, they fail in reproducing the other properties of \astrobj{Sgr dSph} at this time.\\ 

Figure \ref{fig2} also shows that satellites with Plummer radius of 1.2 kpc are destroyed before 5 Gyrs, as we can see in the same figure;  but these systems are the only ones which reproduce the half brightness radius of \astrobj{Sgr dSph}.\\

\begin{figure}
\centering
\includegraphics[width=6.5cm, height=7.5cm]{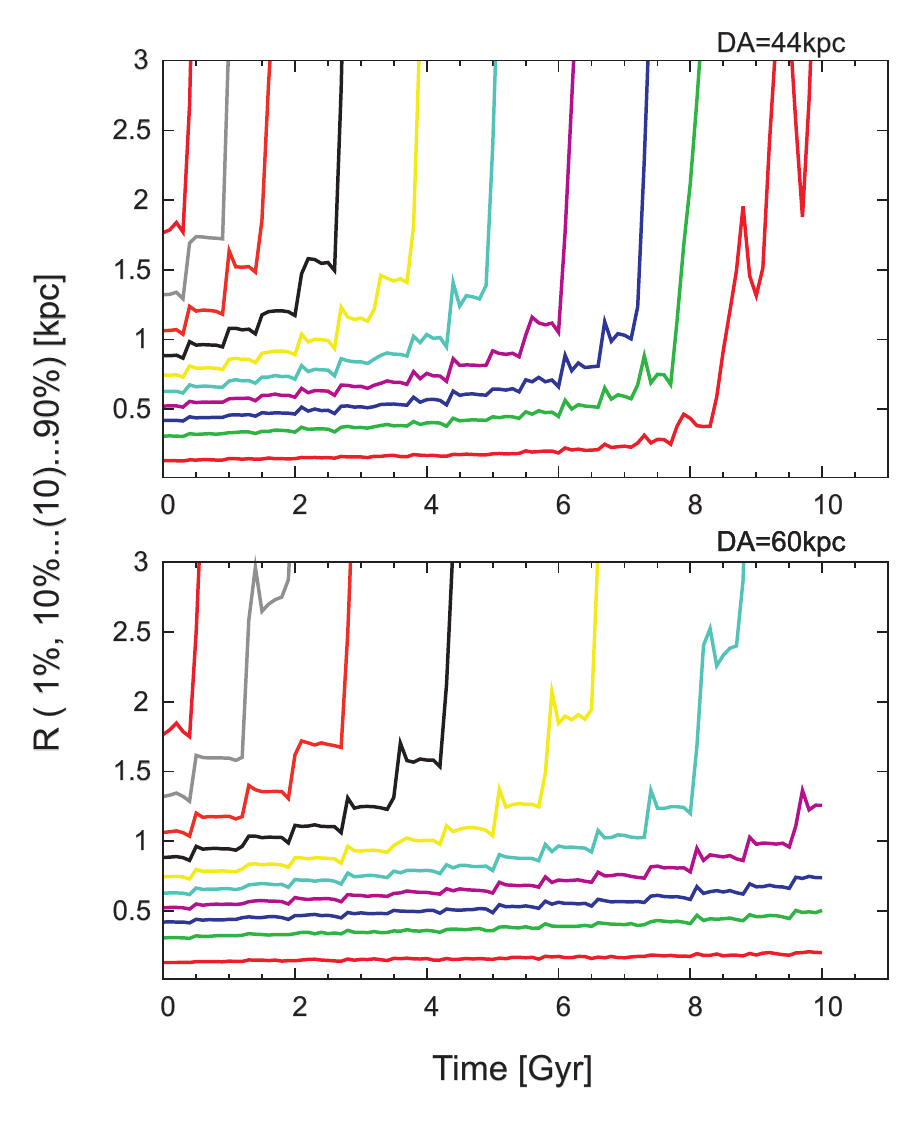}\\
\caption{Evolution of a satellite with initial mass  and Plummer radius of $6\times10^8$ $M_{\odot}$ and 0.6 kpc respectively. This satellite initially is located at a galactocentric distance of: \textit{Top}:  44 kpc. \textit{Bottom}: 60 kpc.}
\label{f3}
\end{figure}

If the satellites are initially located at 60 kpc from the galactic center, their lifetimes are much larger, as we can see in figure \ref{f3} where we show the evolution of the Lagrange radii that contain  from 1\% to 90\% of the initial mass of a galaxy with initial mass of $6\times10^8$ $M_{\odot}$ and a Plummer radius of 0.6 kpc. This galaxy is totally disrupted before 10 Gyrs if it is located at 44 kpc; nevertheless, it can survive for a Hubble time if its initial position is 60 kpc, because 30\% of its initial mass survives  for 10 Gyrs. The same behavior is observed in most of the galaxies in our sample, except for satellites with Plummer radius of 1.2 kpc. Their lifetimes are always lower than 5 Gyrs whenever their initial positions are. \\

\citet{klessen97}; \citet{casas} and \citet{casas12} have argued against the existence of dark matter in dwarf galaxies. In those studies, they obtained large values in  the M/L ratio for spheroidal systems without a dark component.  This  can be explained if it is assumed that  dwarf galaxies are out of equilibrium systems. In this study, we found a remarkable different result:  the M/L ratio for all the simulated satellites without dark matter, is lower than the value measured for \astrobj{Sgr dSph}.  This could be explained if this satellite has a different origin than the other dwarf galaxies of the Milky Way.\\

On the other hand, many observational studies have failed trying to find out evidence of dark matter in dwarf galaxies \citep{farnier, cordier}. Nevertheless, none of our simulations without dark matter could reproduce the measured physical properties of \astrobj{Sgr dSph}. This result suggest  that this satellite, at the beginning of its evolution, should had been immersed in a dark matter halo. Furthermore, the simulations showed in this section point out that its barionic component could not have been a very dense object.

\section{Galaxies immersed in a dark matter halo}
\label{DM}

\begin{figure}
\centering
\includegraphics[width=6.5cm, height=5.5cm]{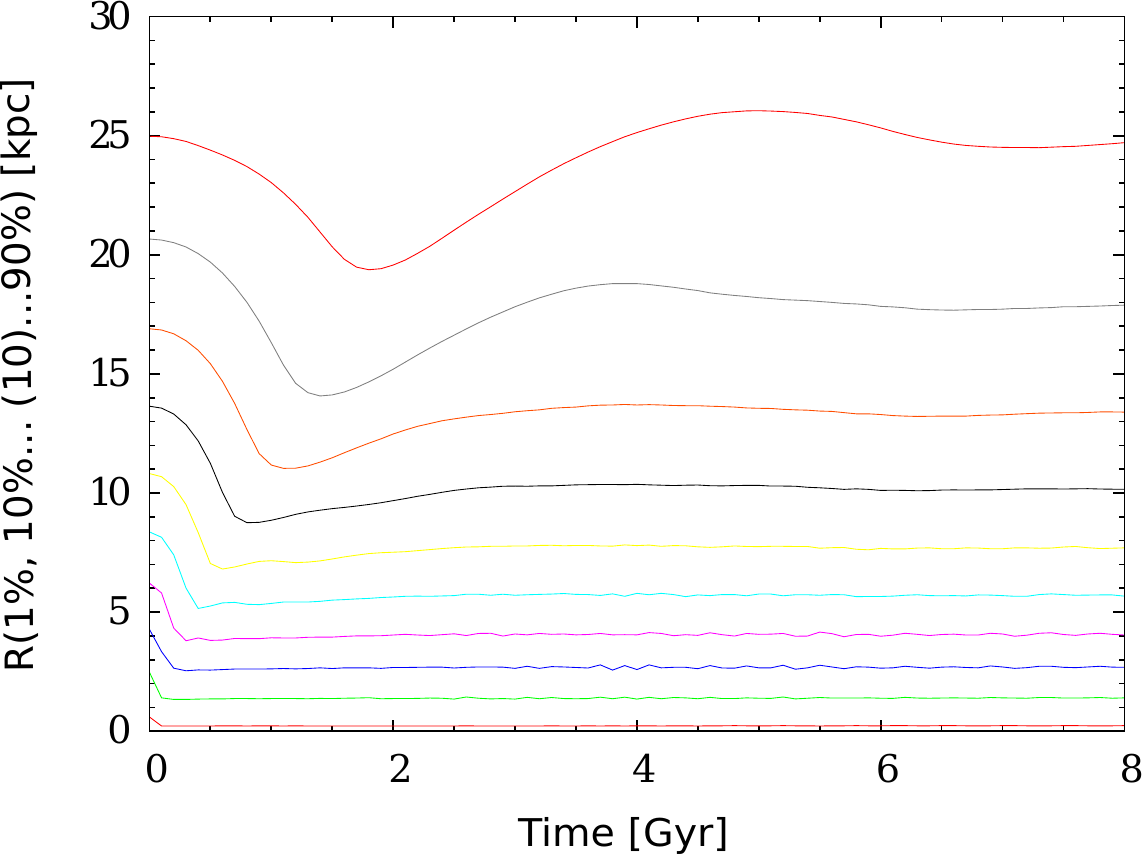} \\
\caption{Relaxation of a system with a dark halo of $8\times 10^8$ $M_{\odot}$,  $r_s= 3$ kpc and concentration of 10.  The barionic component is a satellite of  $7\times 10^8$ $M_{\odot}$ and Plummer radius of 0.6 kpc. Note the collapse of the system during the first 3 Gyr. The system reaches stability after 7 Gyrs. }
\label{halos}
\end{figure}

As we mentioned in section \ref{satellite}, systems composed by barionic and dark matter, were virialized under their self gravity using Gadget 2 \citep{springel05}  until they became stable. Figure \ref{halos} shows this virialization during 8 Gyrs.  \\

The concentration parameter in dark matter halos depends on its mass \citep{lokas01, mashchenko05}, redshift \citep{mashchenko05} and the initial power spectrum of density fluctuations \citep{lokas01}.  The mass concentration (M-c) relation of virialized halos has been determined by various authors using N-body simulations \citep{duffy} or observational data \citep{com, oguri}.\\

Using the M-c relation derived by \citet{duffy}, \citet{com} and \citet{oguri}, we computed the concentration parameter of the dark matter halo that could have had the progenitor of  \astrobj{Sgr dSph} 7 Gyrs ago. We used  $5\times10^8$ $M_{\odot}$ and $1.9\times10^9$ $M_{\odot}$ for the calculations\footnote{Those are the minimum and maximum values used in the simulations.}. Using the M-c relation of \citet{duffy}, we obtained c= 4.52-4.06 for the value of masses respectively. Using the M-c relations of \citet{com} and \citet{oguri} we obtained c= 12.36-10.12 and c= 11.72-10.51 respectively. Taking into account these values, we used for our simulations, concentrations parameters of 4 and 10; furthermore, since numerical simulations suggest that $4\leq c\leq22$ \citep{lokas01}, we made one simulation using c=6.\\

For the simulations of satellites with dark matter, we choose for the barionic component, two types of satellites: not very dense ( $M\leq$ $7\times 10^8$ $M_{\odot}$) and extended  satellites ($r_0= 1.2$ kpc).\\

\begin{figure}
\centering
\includegraphics[width=6 cm, height=5.5cm]{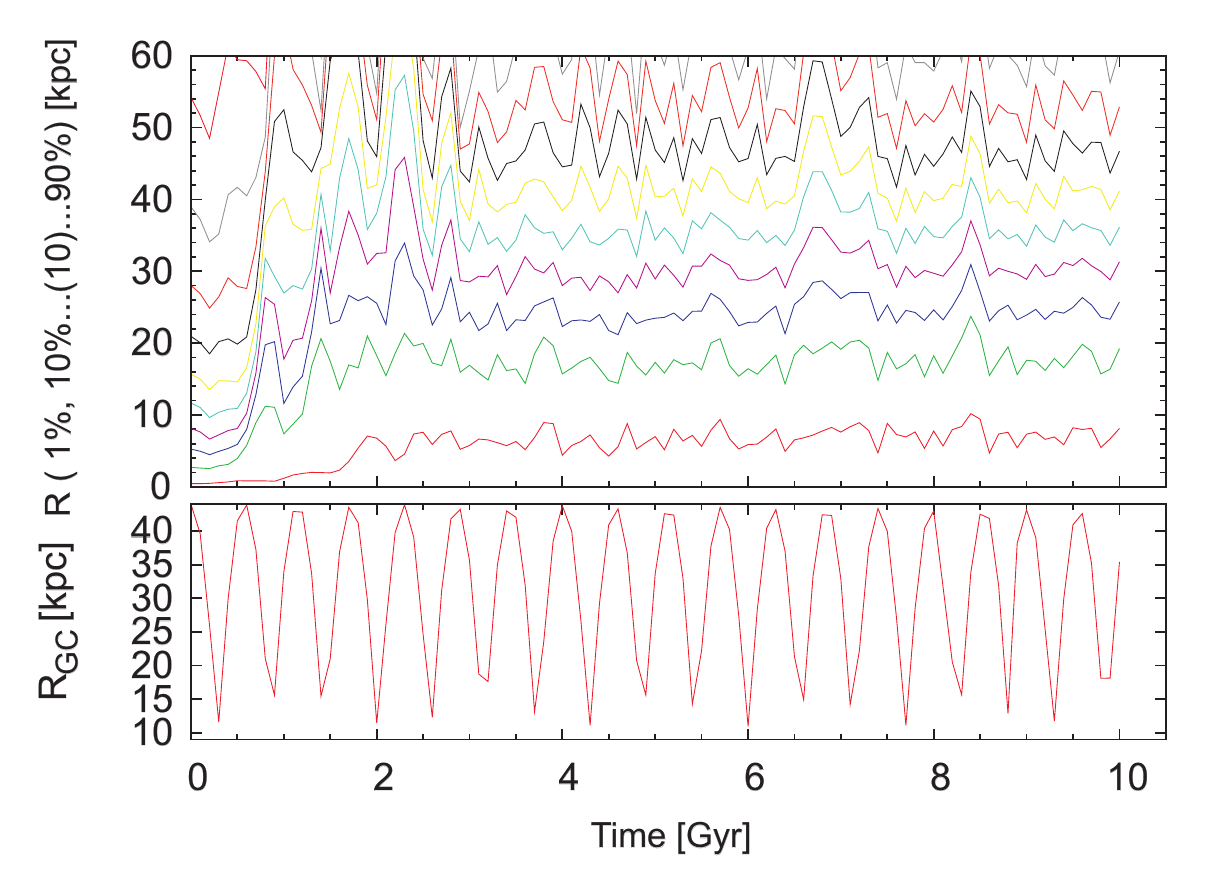} 
\includegraphics[width=6 cm, height=7.5cm]{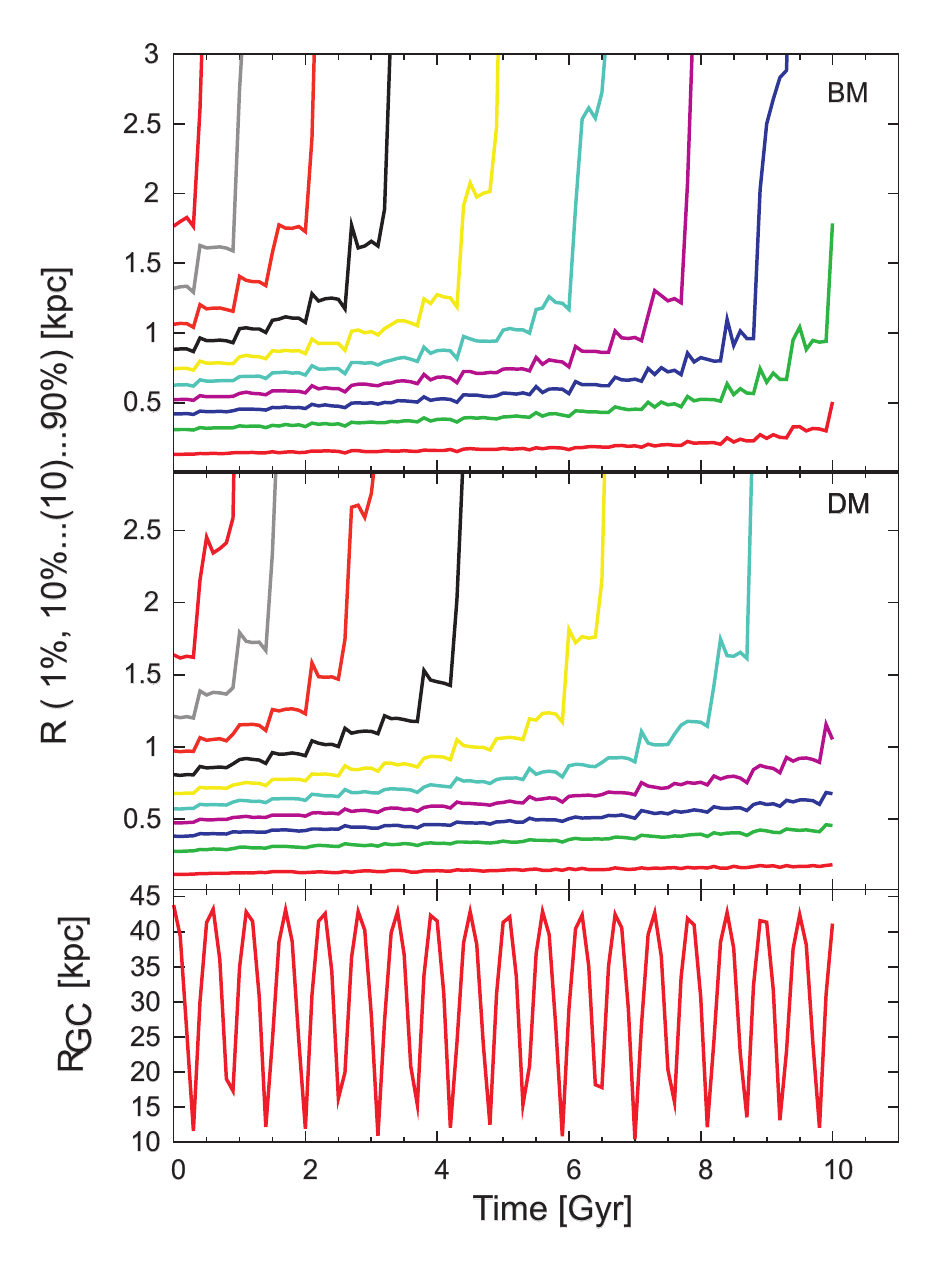}\\
\caption{\textit{Left}:Evolution of a dark halo  with initial mass of $8\times 10^8$ $M_{\odot}$, radius $r_s= 3$ kpc and concentration of 10.  This envelopes a satellite with $7\times 10^8$ $M_{\odot}$ and Plummer radius of 0.6 kpc. The system is located initially at a galactocentric distance of  44 kpc. \textit{Right:} evolution of this satellite with barionic matter and dark matter.}
\label{f4}
\end{figure}

When we studied the dark component of our simulated systems, we observed that the disruption of a dark halo depends on both its initial mass and concentration parameter. As we can see in figure \ref{f4}, in the case of a system whose dark component has a mass of $8\times10^8$ $M_{\odot}$ and concentration of 10, the barionic component begins to be disrupted when the dark halo has lost 44\% of its initial mass. At 2.2 Gyrs of evolution, only remains 10\% of the initial mass of the dark matter halo; while the satellite solely has lost  20\%. In the same figure we show the evolution of the same satellite only with barionic matter. As we expected, the dark component protects the galaxy against tidal disruption. When a system with a dark halo of initial mass of $8\times10^8$ $M_{\odot}$ and concentration parameter of 4 loses 10\% of its mass, the galaxy begins to be disrupted. At 2.8 Gyr of evolution, the dark halo contains only 1\% of its initial mass while the stellar component contains 60\%. \\

We evolve each system immersed in a rigid potential for a time lower than 7 Gyrs; except for the systems with dark matter halos with $9\times10^{6-7}$ $M_{\odot}$  and barionic component of $7\times 10 ^{8}$ $M_{\odot}$, because the fall time given by equation (\ref{fric}) is  larger than 10 Gyr when they are located initially at 44 kpc from the galactic center.  The half light radius, surface brightness, velocity dispersion and mass to light ratio of two galaxies with initial mass of $7\times 10^8$ $M_{\odot}$ and Plummer radii of 0.6 and 1.2 kpc are presented in table \ref{t5}. As in table \ref{t3}, those parameters correspond to the physical properties  observed by a terrestrial observer when each system reproduces the current galactocentric position and radial velocity of \astrobj{Sgr dSph}. Continuous lines represent total disruption of the simulated satellites.\\

\begin{table}
 \centering
  
 \begin{minipage}{140mm}
  \caption{ Physical properties of a galaxy with an initial mass of $7\times10^8$ $M_{\odot}$and Plummer radius of 0.6 and 1.2 kpc when is immersed in different contents of dark matter.  Its initial position is 44 kpc from the galactic centre. Columns have the same meaning as table 3; furthermore, the units of the physical parameters are equal. Note that the half brightness radius of the satellites does not vary with the inclusion of a dark matter halo.}
  \label{t5}
   \vspace{0.5mm}
   \centering
  \begin{tabular}{@{} c c c c c c c c}
  \hline
$r_0$ &\textbf{Halo} &\textbf{t} &\textbf{Mass}  & \textbf{$r_{1/2}$} & $\mu_0$ & $\sigma_0$ & M/L \\ \hline \hline
& M [$M_{\odot}$] \hspace{2mm} $r_s$ [kpc] \hspace{2mm} C  &   & & & & &         \\  
&$8\times10^8$  \hspace{1cm}  3  \hspace{5mm} 10                 & 5.7 & $2.1\times10^8$& 0.14 &$4.71\times10^8$ & 20.20 & 1.95\\ [1.5mm]
 
&$8\times10^8$  \hspace{1cm}  7.5  \hspace{5mm} 4                 & 0.9 & $5.6\times10^8$& 0.18 &$6.36\times10^8$ & 28.77 & 2.30\\ 
&												   & 6.5 & $2.8\times10^8$& 0.16 &$4.1\times10^8$ & 20.31 & 2.05\\  [1.5mm]                                                                                   
0.6&  $9\times10^7$  \hspace{1cm}  7.5  \hspace{5mm} 4          & 0.9 & $5.6\times10^8$& 0.19 &$6.53\times10^7$ & 27.31 & 16.56\\ [1mm]
&                                                                                                   & 8.2 & $1.4\times10^8$& 0.16 &$2.72\times10^7$ & 15 & 16.9\\ [1.5mm]

& $9\times10^6$  \hspace{1cm}  7.5  \hspace{5mm} 4                & 0.9 & $5.6\times10^8$& 0.18 &$7.18\times10^6$ & 27.27 & 184\\ [1mm] 
&												   & 7.7 & $2.1\times10^8$& 0.16 &$3.68\times10^6$ & 16.9 & 160\\ [1mm]\hline\hline

 &$1\times10^7$  \hspace{1cm} 7.5  \hspace{5mm} 4                 & 2.6 & $7\times10^6$  & 0.65 &$1.9\times10^4$ & 21.84 & 22066\\  
 &                                                                                                  & 5.7 & ---& --- &--- & --- & ---\\ [1.5mm]
                                                                                                   
1.2 &$8\times10^7$  \hspace{1cm}  8.3  \hspace{5mm} 6                 & 4.4 & $7\times10^6$& 0.69 &$2.2\times10^3$ & 167 & 600\\[1.5mm]                                                         
                                                                                                   
&$8\times10^8$  \hspace{1cm}  6  \hspace{6mm} 10                 & 4.4 & $7\times10^6$& 0.39 & $7.8\times10^5$ & 93.87 & 500\\ 
                                                                                                    \hline
 
\end{tabular}
\end{minipage}
\end{table}

Simulated galaxies with Plummer radii of 0.6 kpc have a mass consistent with the current range measured in \astrobj{Sgr dSph}, as we can see in table \ref {t5} after 5.7 and 6.5 Gyr of evolution.  Nevertheless, their mass to light ratio is lower than the measured value in \astrobj{Sgr dSph} ; while their surface brightness is higher (these satellites are more luminous than \astrobj{Sgr dSph}). Only  when galaxies are immersed in less massive dark halos, i.e.  $9\times 10^7$ or $9\times 10^6$ $M_{\odot}$, they reproduce most of the physical properties of \astrobj{Sgr dSph}. This result suggests that its initial content of dark matter  could have this range of masses. Nevertheless, when we evolve satellites with initial dark matter halos of $1\times 10^7$ and $8\times10^7$ $M_{\odot}$ and the barionic component with a Plummer radius higher than 0.6 kpc, we can see in table \ref{t5} that they can not survive for a time larger than 5 Gyrs; that is, these simulated galaxies are less stable than \astrobj{Sgr dSph}. This result rejects the fact that the initial content of dark matter of \astrobj{Sgr dSph} could have been of the order of  $10^7$ or $10^6$ $M_{\odot}$.\\

\begin{table}
 \centering
 \begin{minipage}{140mm}
  \caption{Physical properties of a galaxy with an initial mass of $1\times10^9$ $M_{\odot}$ and Plummer radius of 1.2 kpc when is immersed in different contents of dark matter.  Its initial position is 44 kpc from the galactic centre. Third column shows the mass of the barionic matter at time t.}
  \vspace{0.5mm}
   \label{t6}
      \centering
  \begin{tabular}{@{}c c c c c c c}
  \hline
\textbf{Halo} &\textbf{t} &\textbf{Mass} & \textbf{$r_{1/2}$} & $\mu_0$ & $\sigma_0$ & M/L \\ \hline \hline
 M [$M_{\odot}$] \hspace{2mm} $r_s$ [kpc] \hspace{2mm} C  &   & & & & &         \\  
                                                                                                  
 $9\times10^7$  \hspace{1cm} 7.5  \hspace{5mm} 4                 & 0.9 & $1\times10^8$  & 0.45 &$5.8\times10^5$ & 17.86 & 406\\  
                                                                                                   & 5.9 & $---$  & --- &--- & --- & --- \\[1.5mm]  \hline
$5\times10^8$  \hspace{1cm} 6  \hspace{5mm} 10                  & 0.9 & $5\times10^8$  & 0.3 &$9.4\times10^7$ & 22.03 & 6\\  [1.5mm]
\hline
\end{tabular}
\end{minipage}
\end{table}

All of the simulated satellites with Plummer radius of 0.6 kpc survive for 10 Gyrs and as we can see in Figure  \ref{f4}, if the satellite is immersed in a dark matter halo of concentration 10 and initial mass of $8\times 10^8$ $M_{\odot}$, at least 10\% of their initial mass will survive for a Hubble time.\\

When we compare the half brightness radius of satellites with Plummer radius of 0.6 kpc (table \ref{t4}) and 1.2 kpc (tables \ref{t4} and \ref{t5}) we can see that only these galaxies reach a value similar to the observed in \astrobj{Sgr dSph} in some of their stages of evolution. This suggests that at the beginning of the evolution of this dwarf galaxy, it should have been an extended satellite with a Plummer radius larger than 0.5 or 0.6 kpc.\\

\begin{figure}
\centering
\includegraphics[width=6cm, height=7cm]{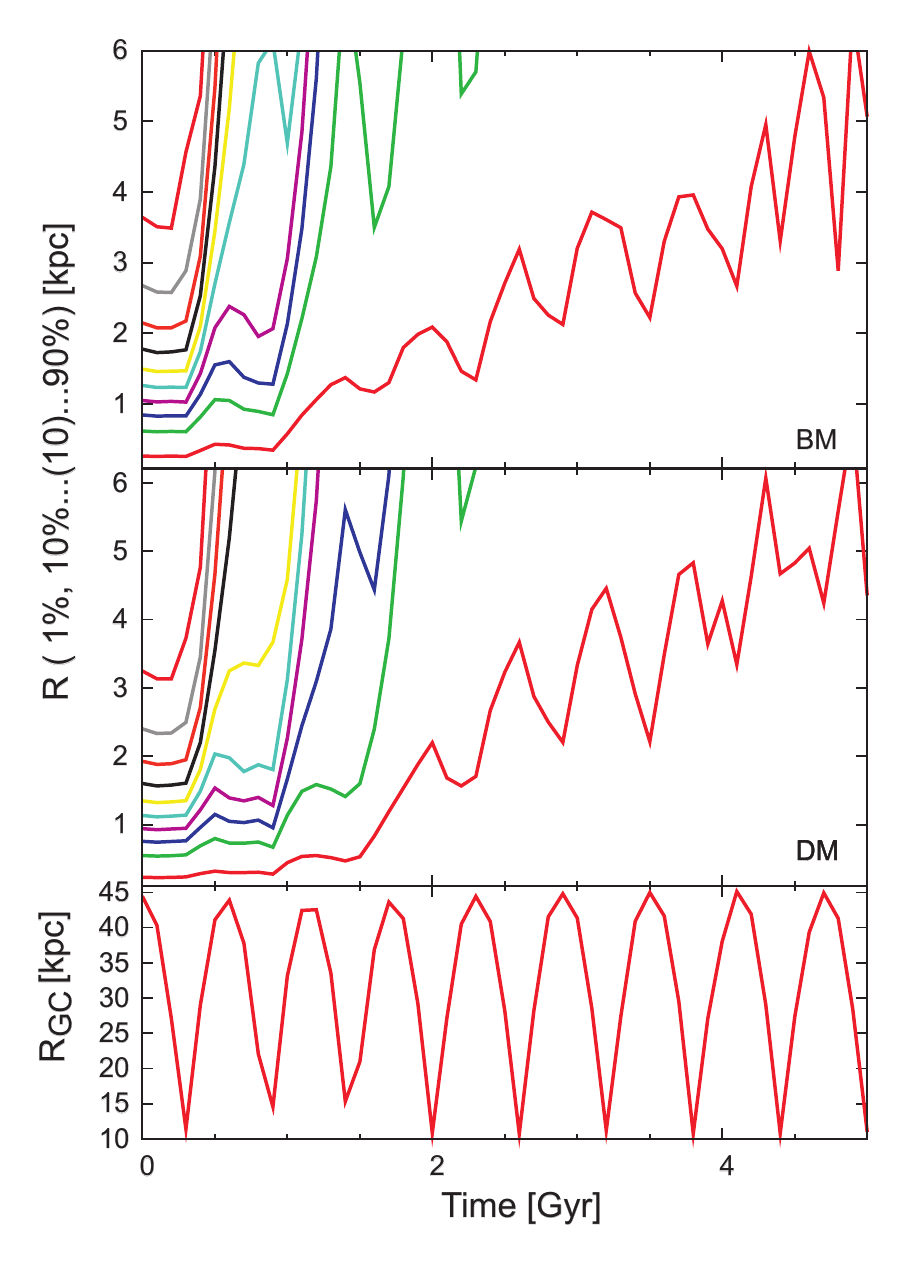} \includegraphics[width=6cm, height=7cm]{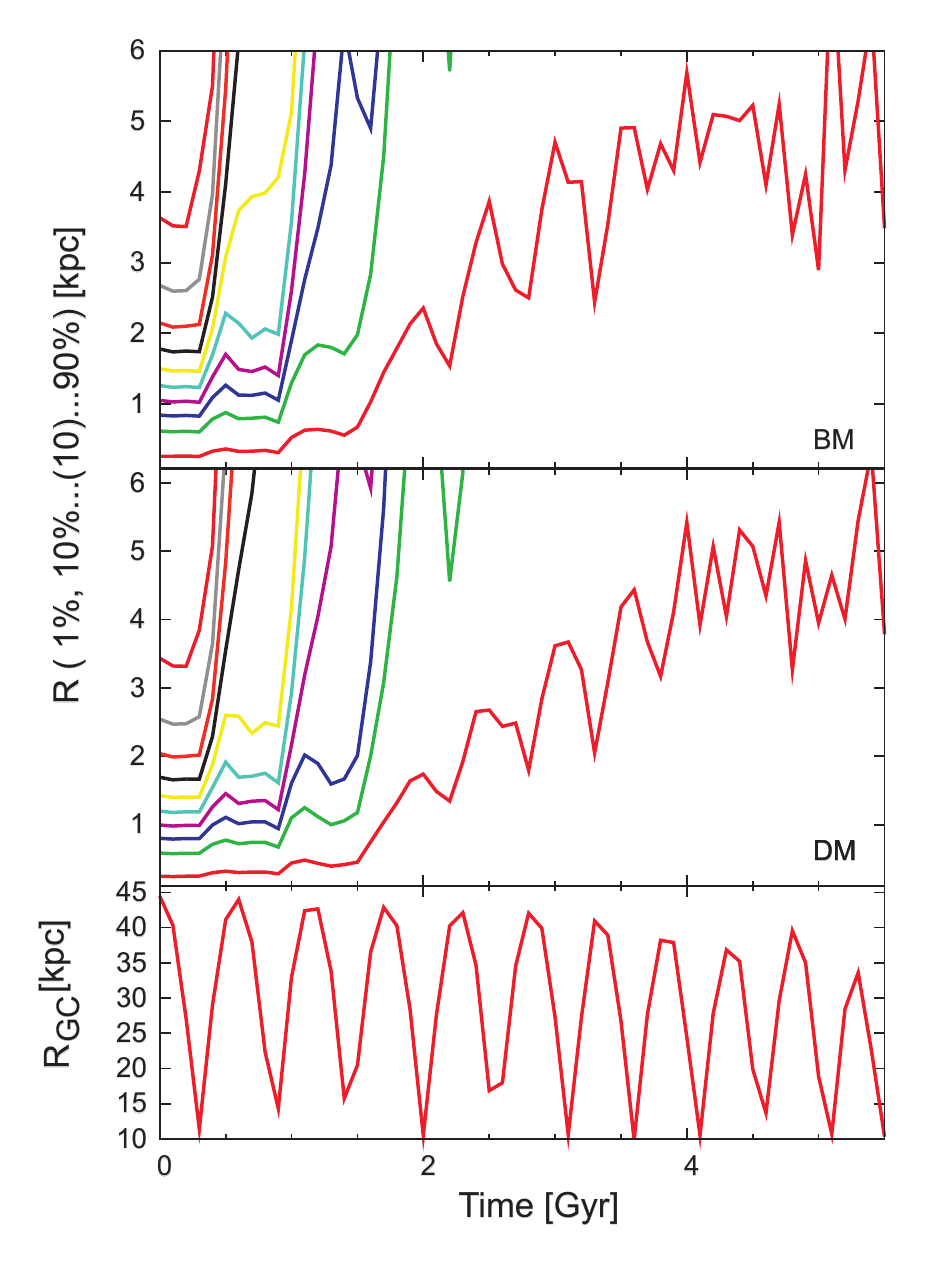}
\caption{Evolution of satellites with Plummer radius of 1.2 kpc and initial mass of: \textit{left:} $7\times10^8$ $M_{\odot}$. \textit{Right:} $1\times10^9$ $M_{\odot}$. Note that there are no difference between a satellite with only barionic matter and the same one immersed in a dark halo.  The satellite of $7\times10^8$ $M_{\odot}$ has a dark halo of  $8\times10^8$ $M_{\odot}$ while the satellite of $1\times10^9$ $M_{\odot}$, another one with $5\times10^8$ $M_{\odot}$. Both have a concentration parameter of 10. Note that these are systems with the maximum allowed mass: $1.5\times10^9$ $M_{\odot}$.  }
\label{f5}
\end{figure}

The evolution  of extended satellites (Plummer radius of 1.2 kpc) immersed in a dark matter halo is presented in figure \ref{f5}. As we noted before, their lifetimes are not higher than 5 Gyrs even when their halos are massive and have a concentration parameter of 10.  Since the radii of these dark matter halos can not be larger, because they could produce some notable effects in the Milky Way, both their initial masses and concentration parameter must be larger than the values used here. It implies that in order to figure out what was the initial dark matter content of \astrobj{Sgr dSph}  using N Body simulations, dynamical friction effects should be taken into account.  The mass of those halos, could possibly had an order of $10^{10}$ or $10^{11}$ $M_{\odot}$. In one of his recent papers, \citet{nierdeste} conclude through measurements of the total luminosity of the main body and tidal tails of \astrobj{Sgr dSph} that the initial mass of its dark matter halo was $\approx 6.3\times10^9$ $M_{\odot}$. This result could be tested using N Body simulations to see if one extended galaxy with this halo can reproduce all the measured physical properties in \astrobj{Sgr dSph}. \\

\section{Conclusions}

We have shown several N body simulations of the evolution of the \astrobj{Sgr dSph} with and without dark matter. In the simulations of satellites with only barionic matter, we found that very dense satellites are more stable than \astrobj{Sgr dSph} because during the entire simulation time, that is, 7 Gyrs, their masses are always larger than the value measured in \astrobj{Sgr dSph}. On the other hand, galaxies with initial masses of $(5-6)\times 10^8$ $M_{\odot}$ reproduced the current bound mass of \astrobj{Sgr dSph} at 6 Gyrs. Extended satellites ($r_0\approx 1.2$ kpc) were the only able to reproduce the half light radius of \astrobj{Sgr dSph}. Despite these results, it was not possible to obtain the entire properties observed in \astrobj{Sgr dSph} by evolving systems with only a barionic component; that is why we conclude that the progenitor of \astrobj{Sgr dSph} should had been immersed in a dark matter halo, as well as the current satellite; despite the fact that many observational studies have failed in finding observational evidence of dark matter in dwarf galaxies.\\

Simulations of systems with dark and barionic matter suggest that  the progenitor of \astrobj{Sgr dSph} had to be an extended galaxy ($r_0 \geq 0.6$ kpc) in order to reproduce the observed half light mass radius; furthermore, the initial mass of its  dark matter halo  had to be larger than $10^{8}$ $M_{\odot}$. 

\section*{Acknowledgments}

We acknowledge to Sergey Mashchenko for his help in implementing his initial conditions generator for dark halos.  We would like to thank Leonardo Casta\~neda and Irapuan Rodrigues  for their useful comments as well as the anonymous referee for improving this paper with very important suggestions.


\begin{thebibliography}{00}

\bibitem[\protect\citeauthoryear{Aarseth et al.}{1974}]{aarseth74} Aarseth, S.J., Hénon, M., Wielen, R.: 1974, \textit{A\&A}  \textbf{37}, 183
\bibitem[\protect\citeauthoryear{Aarseth}{2003}]{aarseth03}  Aarseth, S. J.: 2003, \textit{Gravitational N-Body simulations}, Cambridge University Press. 

\bibitem[\protect\citeauthoryear{Alard}{1996}]{alard96} Alard, C.: 1996, \textit{ApJ} \textbf{458}, L17

\bibitem[\protect\citeauthoryear{Belokurov et al.}{2006}]{belokurov06} Belokurov, V., Zucker, D.B., Evans, N.W., Gilmore, G., Vidrih, S., Bramich, D.M., Newberg, H.J., Wyse, R.F.G., Irwin, M.J., Fellhauer, M., Hewett, P.C., Walton, N.A., Wilkinson, M.I., Cole, N., Yanny, B., Rockost, C.M., Beers, T.C., Bell, E.F., Brinkmann, J., Ivezic, Z., Lupton,R.:  2006, \textit{ApJ} \textbf{642}, L137

\bibitem[\protect\citeauthoryear{Binney and Tremaine}{1994}]{binney}  Binney, J., Tremaine, S.: 1994, \textit{Galactic Dynamics}, Princeton University Press. 

\bibitem[\protect\citeauthoryear{Casas and Kroupa}{2011}]{casas} Casas, R.A., Kroupa, P.: 2011, \textit{RMxAC} \textbf{40}, 48C
\bibitem[\protect\citeauthoryear{Casas et al.}{2012}]{casas12} Casas, R.A., Arias, V., Pena Ram\'irez, K., Kroupa, P.: 2012, \textit{MNRAS} \textbf{424}, 1941

\bibitem[\protect\citeauthoryear{Comerford and Natarajan}{2007}]{com} Comerford, J.M., Natarajan, P.: 2007, \textit{MNRAS} \textbf{379}, 190 


\bibitem[\protect\citeauthoryear{Cordier et al.}{2004}]{cordier} Cordier, B., Atti\'e, D., Cass\'e, M., Paul, J., Schanne, S., Sizun, P., Jean, P., Roques, J.P., Vedrenne, G.: 2004, \textit{ESASP} \textbf{552}, 581C

\bibitem[\protect\citeauthoryear{Dinescu et al.}{2002}]{dinescu02} Dinescu D. I.,  Majewski S. R., Girard, T.M., M\'endez, R.A., Sandage, A., Siegel, M.H., Kinkel, W.E., Subasavage, J.P., Ostheimer, J.:  2002, \textit{ApJ} \textbf{575}, L67

\bibitem[\protect\citeauthoryear{Duffy et al.}{2008}]{duffy}
Duffy, A.R., Schaye, J., Kay, S.T., Dalla Veccia, C.: 2008, \textit{MNRAS} \textbf{390}, L64

 
\bibitem[\protect\citeauthoryear{Farnier}{2007}]{farnier} Farnier C., H.E.S.S. collaboration: 2007,  \textit{SF2A} ,168

\bibitem[\protect\citeauthoryear{Fellhauer et al.}{2006}]{fellhauer06} Fellhauer M., Belokurov V., Evans, N.W., Wilkinson, M.I., Zucker, D.B., Gilmore, G., Irwin, M.J., Bramich, D.M., Vidrih, S., Wyse, R.F.G., Beers, T.C., Brinkmann, J.:  2006, \textit{ApJ} \textbf{651}, 167

\bibitem[\protect\citeauthoryear{Giuffrida et al.}{2010}]{giuffrida09} Giuffrida G., Sbordone L., Zaggia, S., Marconi, G., Bonifacio, P., Izzo, C., Szeifert, T., Buonanno, R.:   2010, \textit{A\& A} \textbf{513A}, 62G

\bibitem[\protect\citeauthoryear{G\'omez-Flechoso et al.}{1999}]{gomez99} G\'omez-Flechoso M.A. , Fux R., Martinet L.: 1999, \textit{A\&A} \textbf{347}, 77

\bibitem[\protect\citeauthoryear{Helmi}{2003}]{helmi03} Helmi, A., Zeljko, I., Prada, F., Pentericci, L., Rockosi, C.M., Schneider, D.P., Grebel, E.K., Harbeck,D., Lupton, R.H., Gunn, J.E., Knapp, G.R., Strauss, M.A., Brinkmann, J.: 2003, \textit{ApJ} \textbf{586}, 195H

\bibitem[\protect\citeauthoryear{Helmi}{2004}]{helmi04} Helmi, A.: 2004, \textit{ApJ} \textbf{610}, L97

\bibitem[\protect\citeauthoryear{Hernquist}{1990}]{hernquist90} Hernquist, L.: 1990, \textit{ApJ} \textbf{356}, 359

\bibitem[\protect\citeauthoryear{Ibata, Gilmore and Irwin}{1995}]{ibata95}  Ibata, R. A., Gilmore, G., Irwin,  M. J.: 1995, \textit{MNRAS} \textbf{277}, 781

\bibitem[\protect\citeauthoryear{Ibata et al.}{1997}]{ibata97} Ibata, R. A., Wyse, R. F.G., Gilmore, G., Irwin, M. J., Suntzeff, N. B.: 1997, \textit{AJ} \textbf{113}, 634I

\bibitem[\protect\citeauthoryear{Ibata and Lewis}{1998}]{ibata98} Ibata, R.A., Lewis, G.F.: 1998, \textit{ApJ} \textbf{500}, 575

\bibitem[\protect\citeauthoryear{Ibata}{1999}]{ibata99} Ibata, R.A.: 1999, \textit{IAUS} \textbf{186}, 39I

\bibitem[\protect\citeauthoryear{Ibata et al.}{2001}]{ibata01} Ibata, R., Lewis, G. F., Irwin, M., Totten, E., and Quinn, T.: 2001, \textit{ApJ} \textbf{551}, 294



\bibitem[\protect\citeauthoryear{Irwin and Hatzidimitriou}{1995}]{irwin95} Irwin, M. and Hatzidimitriou, D.: 1995, \textit{MNRAS} \textbf{277}, 1354

\bibitem[\protect\citeauthoryear{Irwin et al.}{1996}]{irwin96} Irwin M., Ibata R., Gilmore G., Wyse R., Suntzeff N.: 1996, \textit{ASPC} \textbf{92}, 84

\bibitem[\protect\citeauthoryear{Ing-Guey and Binney}{2000}]{jiang00} Ing-Guey, J. and Binney, J.: 2000, \textit{MNRAS} \textbf{314}, 478 

\bibitem[\protect\citeauthoryear{Johnston et al.}{1995}]{johnston95} Johnston, K. V., Spergel, D. N., Hernquist, L.: 1995, \textit{ApJ} \textbf{451}, 598

\bibitem[\protect\citeauthoryear{Johnston et al.}{1999}]{johnston99} Johnston, K.V., Majewsky, S.R., Siegel, M.H., Reid, I.N., Kunkel, W.E.: 1999, \textit{AJ} \textbf{118}, 1719


\bibitem[\protect\citeauthoryear{Keller et al.}{2008}]{keller08}  Keller, S. C., Murphy, S., Prior, S., DaCosta, G., Schmidt, B.: 2008, \textit{ApJ} \textbf{678}, 851

\bibitem[\protect\citeauthoryear{Klessen and Kroupa}{1998}]{klessen97} Klessen, R.S. and Kroupa, P.: 1998, \textit{ApJ} \textbf{498}, 143

\bibitem[\protect\citeauthoryear{Klimentowski et al.}{2009}]{klimentowski09} Klimentowski , J.,  Lokas, E. L., Kazantzidis, S., Mayer, L., Mamon, G.A.:  2009, \textit{MNRAS} \textbf{397}, 2015K 

\bibitem[\protect\citeauthoryear{Law et al.}{2003}]{law03}  Law, D. R., Majewski, S. R., Johnston, K.,  Skrutskie, M. F.: 2003, \textit{ASP Conference Series}
\textbf{327}, 239L

\bibitem[\protect\citeauthoryear{Law et al.}{2005}]{law05}  Law, D. R., Johnston, K. V., Majewski, S. R.: 2005, \textit{ApJ} \textbf{679}, 807

\bibitem[\protect\citeauthoryear{Lokas and Mamon}{2010}]{lokas01} Lokas, E.L., and Mamon, G.A.: 2001, \textit{MNRAS} \textbf{321}, 155 

\bibitem[\protect\citeauthoryear{Law and Majewski}{2010}]{law10} Law, D.R., and Majewski, S.R.: 2010, \textit{ApJ} \textbf{718}, 1128 

\bibitem[\protect\citeauthoryear{Majewski et al.}{2003}]{majewsky03a}  Majewski, S. R., Skrutskie, M.F., Weinberg, M. D.,  Ostheimer, J. C.: 2003, \textit{ApJ}  \textbf{599}, 1082M


\bibitem[\protect\citeauthoryear{Majewski et al.}{2004}]{majewsky03c}  Majewski, S. R., Law, D., Johnston, K.,  Skrutskie, M.F., Weinberg, M. D., 2004, \textit{ASP Conference Series} \textbf{220}, 189

\bibitem[\protect\citeauthoryear{Mart\'inez-Delgado et al.}{2004}]{martinez04} Mart\'inez-Delgado, D., G\'omez-Flechoso, A. M., Aparicio, A., Carrera, R.: 2004, \textit{ApJ} \textbf{601}, 242

\bibitem[\protect\citeauthoryear{Mart\'inez-Delgado et al.}{2001}]{martinez01} Martinez-Delgado, D., and Aparicio, A.: 2001, \textit{ASPC} \textbf{245}, 352M

\bibitem[\protect\citeauthoryear{Mashchenko and Sills}{2005}]{mashchenko05} Mashchenko, S., and Sills, A.: 2005, \textit{ApJ} \textbf{619}, 243

\bibitem[\protect\citeauthoryear{Mateo  et al.}{1998}]{mateo98} Mateo, M., Olszewski, E. W., Morrison, H. L., 1998, \textit{ApJ} \textbf{508}, L55

\bibitem[\protect\citeauthoryear{Miyamoto and Nagai}{1975}]{miyamoto75} Miyamoto, M., and Nagai, R.: 1975, \textit{PASJ} \textbf{27}, 533

\bibitem[\protect\citeauthoryear{Newberg et al.}{2002}]{newberg02} Newberg, H., Yanny, B., et al. 2002, ApJ, 569, 245

\bibitem[\protect\citeauthoryear{Newberg et al.}{2003}]{newberg03}  Newberg, H.J., Yanni, B., Grebel, E.K., Hennessy, G., Zeljko,I., Mart\'inez-Delgado, D., Odenkirchen, M., Rix, H.W., Brinkmann, J., Lamb, D.Q., Schneider, D.P., York, D.G.:  2003, \textit{ApJ}  \textbf{596}, L191

\bibitem[\protect\citeauthoryear{Nierdeste-Ostholt et al.}{2010}]{nierdeste} Niederste-Ostholt, M., Belokurov, V., Evans,  N.W., Pe\~narrubia, J.: 2010, \textit{ApJ} \textbf{712}, 516 

\bibitem[\protect\citeauthoryear{Oguri et al.}{2009}]{oguri} Oguri, M., Hennawi, J.F., Gladders, M.D., Dahle, H., Natarajan, P., Dalal, N., Koester, B.P., Sharon, K., Bayliss, M.: 2009, \textit{MNRAS} \textbf{699}, 1038


\bibitem[\protect\citeauthoryear{Pe\~narrubia et al.}{2010}]{penarrubia10} Pe\~narrubia, J., Belokurov, V.,  Evans, N.W., Mart\'inez-Delgado, D., Gilmore, G., Irwin, M., Nierdeste-Ostholt, M., Zucker, D.B.:  2010, \textit{MNRAS} \textbf{408}, 26P

\bibitem[\protect\citeauthoryear{Plummer}{1911}]{plummer11} Plummer, M.A.: 1911, \textit{MNRAS} \textbf{71}, 460P

\bibitem[\protect\citeauthoryear{Springel}{2005}]{springel05}  Springel, V.: 2005, \textit{MNRAS} \textbf{364}, 1105

\bibitem[\protect\citeauthoryear{Watkins et al.}{2009}]{watkins09} Watkins, L.L. , Evans,  N.W., Belokurov, V., Smith, M.C., Hewett, P.C., Bramich, D.M., Gilmore, G.F., Irwin, M.J., Vidrih, S., Wyrzykowski, L., Zucker, D.B.:  2009, \textit{MNRAS} \textbf{398}, 1757

\bibitem[\protect\citeauthoryear{Yanny et al.}{2009}]{yanny09} Yanny, B., Newberg, H.J., Johnston, J.A., Lee, Y.S.,  Beers, T.C., Bizyaev, D., Brewington, H., Re Fiorentin, P., Harding, P., Malanushenko, E., Malanushenko, V., Oravetz,D., Pan,K., Simmons, A., Snedden,S.:   2009, \textit{ApJ} \textbf{700}, 1282

\end{thebibliography}
\end{document}